\documentclass[10pt,conference]{IEEEtran} 
\IEEEoverridecommandlockouts

\usepackage{todonotes}

\usepackage{amssymb}
\usepackage{amsmath, amsfonts}

\usepackage{graphicx}
\usepackage{textcomp}
\usepackage{xcolor}
\usepackage{hyperref}

\usepackage{xspace}
\usepackage{paralist}
\usepackage{multirow}
\usepackage{subcaption}
\usepackage{tikz}
\usepackage{courier}
\usepackage{listings} 
\usepackage{array}
\usepackage{graphics}
\usepackage{lipsum}
\usepackage{wrapfig}
\usepackage{arydshln}
\usepackage{url}
\usepackage{alltt}
\usepackage{comment}
\usepackage{booktabs} 
\usepackage[numbers]{natbib}

\usepackage[capitalise]{cleveref}

\newcommand*\circled[2][fill=black]{\tikz[baseline=(char.base)]{
    \footnotesize
    \node[shape=circle, #1, inner sep=1pt] (char) {\textcolor{white}{#2}};}}
\newcommand{\etal}{\emph{et al.}\xspace}

\NewDocumentCommand{\rangeet}
{ mO{} }{\textcolor{blue}{\textsuperscript{\textit{rangeet}}\textsf{\textbf{\small[#1]}}}}
\newcommand{\tdd}{TDD-Bench-Verified\xspace}
\newcommand{\atdd}{Auto-TDD\xspace}

\usepackage{cuted}

\definecolor{Gray}{gray}{0.3}
\tikzstyle{mybox} = [draw=black, very thick, rectangle, rounded corners, inner ysep=5pt, inner xsep=5pt, fill=gray!20]

\newcommand{\findings}[2]{
    \smallskip
    \noindent
    \begin{tikzpicture}
        \node [mybox] (box){%
        \centering
        \begin{minipage}{.95\columnwidth}
        \fontsize{8.8}{10}\selectfont
        \textbf{Finding #1}. #2
        \end{minipage}
        };
    \end{tikzpicture}%
}

\begin{document}
\title{TDD-Bench Verified: Can LLMs Generate Tests for Issues Before They Get Resolved?}


\author{Toufique Ahmed, Martin Hirzel, Rangeet Pan, Avraham Shinnar, and Saurabh Sinha\\ IBM Research}

\maketitle

\begin{figure*}[!htp]
    \centering
    \includegraphics[width=\linewidth]{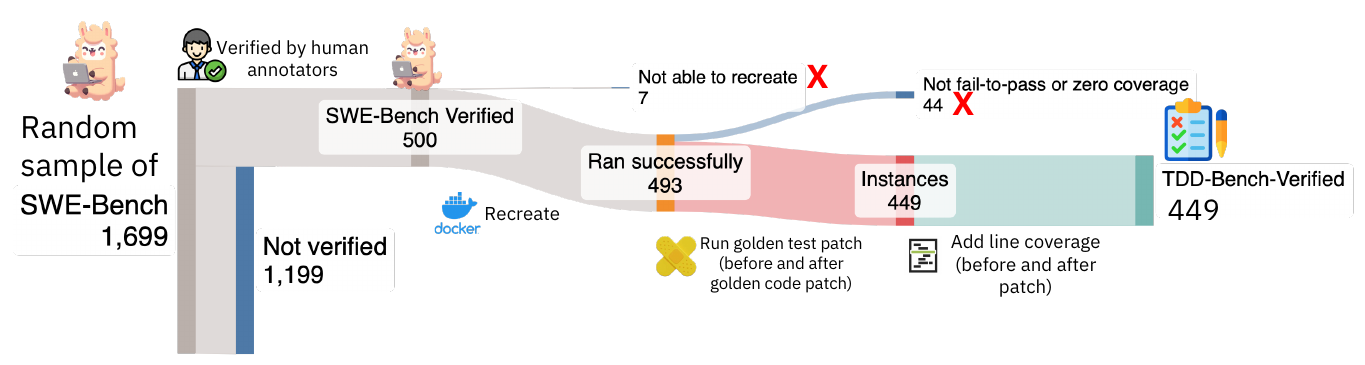}
    \vspace{-.2cm}
    \caption{Overall flow of TDD-bench dataset filtering starting from SWE-bench verified}
    \label{fig:tdd-flow}
    \vspace{-.3cm}
\end{figure*}

\begin{abstract}
Test-driven development (TDD) is the practice of writing tests first
and coding later, and the proponents of TDD expound its numerous benefits.
For instance, given an issue on a source code repository, tests can
clarify the desired behavior among stake-holders before anyone writes
code for the agreed-upon fix.
Although there has been a lot of work on automated test
generation for the practice ``write code first, test later'', there
has been little such automation for TDD.
Ideally, tests for TDD should be fail-to-pass (i.e., fail before the issue
is resolved and pass after) and have good adequacy with respect to covering the code changed
during issue resolution.
This paper introduces TDD-Bench Verified, a high-quality
benchmark suite of 449 issues mined from real-world GitHub code
repositories.
The benchmark's evaluation harness runs only relevant tests in
isolation for simple yet accurate coverage measurements, and the
benchmark's dataset is filtered both by human judges and by execution
in the harness.
This paper also presents Auto-TDD, an LLM-based solution that takes as
input an issue description and a codebase (prior to issue resolution) and returns as
output a test that can be used to validate the changes made for resolving the issue.
Our evaluation shows that Auto-TDD yields a better fail-to-pass rate
than the strongest prior work while also yielding high coverage
adequacy.
Overall, we hope that this work helps make developers more productive
at resolving issues while simultaneously leading to more robust fixes.
\end{abstract}
\begin{IEEEkeywords}
test-driven development, test generation, LLMs, benchmarks
\end{IEEEkeywords}





\section{Introduction}
\label{sec:intro}

Benchmarks can inspire technological progress, but benchmarks for
automated software engineering tasks lag behind the increased adoption
of stronger and stronger large language models~(LLMs).
To be meaningful, benchmarks need to be realistic and measurable while
also being challenging.
For instance, the HumanEval benchmark~\cite{chen2021evaluating} is
measurable and was, initially, challenging for LLMs, but more recently
lost popularity when LLM performance on it became saturated.

One important software engineering task that can benefit greatly
from an up-to-date benchmark is
Test-driven development, or \emph{TDD}~\cite{beck_2002}.
TDD is the
practice of ``test first, write code later'', where a software
developer writes tests before writing corresponding code.
This means the tests initially fail, and, if everything goes right,
they pass after applying the code changes.
Compared to the common practice of ``write first, test later'', TDD
makes requirements clearer, enhances confidence in the code once
written, and leads to tests that emphasize the interface over
implementation details.
For example, up-front tests can clarify the desired behavior between
the interested parties for an issue on a source code repository,
including the project maintainer, the user opening the issue, and the
developer submitting a pull request to close it.
Subsequently, the same tests can serve as acceptance criteria for the pull request
once the code is written.

This paper contributes a new benchmark, \tdd.
This new benchmark is derived from SWE-bench~\cite{jimenezswe}, a
dataset for issue resolution comprising 2,294~issues mined from
12~popular Python GitHub repositories.
Deriving \tdd involved two modifications: filtering for high-quality
instances and evaluating test generation instead of issue resolution.
For filtering, it reuses an extensive human annotation campaign done by
OpenAI~\cite{chowdhury_et_al_2024} to avoid underspecified issues,
overly-specific tests, and flaky test environments.
While the OpenAI annotation campaign filtered the dataset down to 500
issues, some of the remaining issues were still problematic.
Therefore, we applied additional automated filters, resulting in an
even higher-quality subset of 449 issues.
Each issue yields one \emph{instance}
\mbox{$x=\langle d_\mathrm{issue},c_\mathrm{old}\rangle$} comprising a
natural-language issue description $d_\mathrm{issue}$ together with
the original version of a \mbox{codebase $c_\mathrm{old}$} right
before the issue was addressed.

A \emph{solution} to \tdd consists of a function $\mathit{genTests}$
that takes an instance $x$ and returns a set of
\mbox{tests $y=\mathit{genTests}(x)$}.
\tdd provides an \emph{evaluation harness} that uses various testing tools in a
containerized environment to implement an evaluation metric
$\mathit{tddScore}$ that a solution $\mathit{genTests}$ tries to
maximize.
While the solution $\mathit{genTests}$ has access to $c_\mathrm{old}$
only, the evaluation metric also uses the hidden golden new
\mbox{code $\hat{c}_\mathrm{new}$} right after the issue was fixed.
The metric $\mathit{tddScore}$ combines two factors.
First, $\mathit{failToPass(x, y)}$ is a binary correctness
metric that checks whether the tests $y$ fail on the old code
$c_\mathrm{old}$ and pass on~$\hat{c}_\mathrm{new}$.
Satisfying the $\mathit{failToPass}$ criterion is necessary but not
sufficient for a good test suite.
Therefore, the second factor, $\mathit{adequacy}(x, y)$, measures
adequacy of tests $y$ with respect to instance $x$ via
coverage on the old and new code.

Because writing tests up-front is tedious for humans, recent work has
started automating that task using LLMs.
\textsc{Libro}~\cite{kang2023large} prompts Codex~\cite{chen2021evaluating} with
Java issues from Defects4J~\cite{just2014defects4j} and achieves 33\%
success rate in creating fail-to-pass tests for 750 issues, prompting
the LLM 50~times for each issue (pass@50); when considering only one generation
per issue (pass@1), its success rate drops to 19.9\%.
Plein et al.~\cite{plein_et_al_2024} also prompt LLMs with Defects4J
issues, but generate fail-to-pass tests
in only 6\% of cases using ChatGPT. 
M{\"u}ndler et al.\ \cite{mundler2024code} introduce a benchmark,
SWT-bench, that is similar to \tdd in that it evaluates how well a
solution can generate fail-to-pass tests from issue descriptions.
However, SWT-bench applies less rigorous quality filters than \tdd, and it measures
coverage in a more round-about way by first running more tests than
just the submitted ones and then subtracting them back out.
The paper that introduces SWT-bench \cite{mundler2024code} also experiments
with various solutions, but only evaluates them on a subset of 276 single-file
issues called SWT-bench Lite instead of the full 1,983 instances of SWT-Bench.
Their best solution, \mbox{SWE-Agent+}, is derived from
SWE-Agent~\cite{sweagent2}, and using \mbox{GPT-4} achieves a
fail-to-pass rate of 19.2\% on SWT-bench Lite.
Similarly, \textsc{Libro}, while it performs better on Defects4J
issues with multiple trials per issue, achieves a fail-to-pass rate of
only 15.2\% on SWT-Bench Lite.
These numbers indicate that generating tests from issues is challenging even
for the latest frontier models with the latest agents.

In addition to our new benchmark, this paper also introduces
\atdd, our new solution that achieves 21.7\%
fail-to-pass rate on SWT-bench Lite, outperforming the best prior
solution \mbox{SWE-Agent+}~\cite{mundler2024code} while being simpler.
\atdd accomplishes this strong performance through a combination
of prompting and symbolic techniques.
It decomposes the task into a 3-step pipeline of LLM calls, making the
problem solved by each step simpler while also being more predictable
end-to-end than a fully dynamic agent.
It is designed to work not only with frontier models but also with
open-source models, by using few-shot prompts to help models better
understand the task and format at hand.
\atdd uses symbolic techniques to gather and render the right code
context to use as LLM input, and has a simple LLM output format
designed to better match the pre-training distribution of the model.
In addition, it uses symbolic techniques to repair mistakes in
LLM-generated code before submitting a test.

Most evaluations in this paper are based on the 449 instances in \tdd,
except when a direct comparison to \mbox{SWE-Agent+} necessitated the
use of the 276 instances used by their paper~\cite{mundler2024code}.
The experiments used three LLMs (\mbox{llama-3.1-70b},
\mbox{mistral-large}, and GPT-4o).
The best-performing LLM was GPT-4o with a fail-to-pass rate of 23.6\%
on \tdd.
We conducted various ablation experiments finding, among other things,
that LLM-based test file selection was crucial.
While our primary focus was TDD, we also experimented with a variant
of the benchmark and solution for a ``write code first, test later''
scenario.
In terms of adequacy, we found bimodal results.
When model-generated tests were fail-to-pass, their coverage was above
90\% (similar to human-written tests), but other model-generated tests
had a coverage of less than 60\%.

The contributions of this paper are:

\begin{itemize}
    \item A new benchmark, \tdd, for the practice of ``test first, write code later'',  that evaluates the correctness and adequacy of tests generated from issue descriptions in real-world software projects, available at
    \textcolor{blue}{\url{https://github.com/IBM/TDD-Bench-Verified}}.

    \item A new technique, \atdd, that dramatically improves the state of the art for generating tests before the code-to-be-tested is written, by combining new effective LLM prompting techniques with symbolic techniques.
    
    \item An evaluation of \atdd on \tdd, along with a qualitative and quantitative exploration of related aspects around test adequacy, human-written tests, and the alternative practice of ``code first, test later''.
\end{itemize}

We hope that \tdd and \atdd will inspire improvements in automated
test-driven development, improve developer productivity, and
ultimately lead to more robust software.

\section{\tdd Benchmark}
\label{sec:tddbench}

This section introduces \tdd, our new benchmark that requires
generating tests given only an issue description and an old
version of the code, but without access to the new code to be tested.

\subsection{SWE-Bench}\label{sec:swe_bench_verified}

\tdd builds upon SWE-bench~\cite{jimenezswe}.
SWE-bench was mined from GitHub pull requests (PRs) that resolved
issues.
Each SWE-bench instance is a pair
\mbox{$x=\langle d_\mathrm{issue},c_\mathrm{old}\rangle$}
of an issue description $d_\mathrm{issue}$ alongside the old version
$c_\mathrm{old}$ of the code just before the PR.
Furthermore, the SWE-bench evaluation harness uses a
set of golden tests $\hat{y}$ and golden new code $\hat{c}_\mathrm{new}$ mined from the same PR.

A solution to SWE-bench is a function $\mathit{genCode}$ that takes an
instance $x$ and returns a new version
\mbox{$c_\mathrm{new}=\mathit{genCode}(x)$} of the code.
The golden new code $\hat{c}_\mathrm{new}$ and golden tests $\hat{y}$
in the mined PR are hidden from the solution $\mathit{genCode}$, which
only has access to~$x$.
SWE-bench evaluates a solution $\mathit{genCode}$ by the sum, over all
instances $x$, of the pass criterion
\mbox{$\mathit{pass}(x,c_\mathrm{new})=I\big(\mathit{fail}\notin\mathit{runTests}(\hat{y},c_\mathrm{new})\big)$}.
(Here, $I(p)$ is the indicator function that returns 1 if
predicate $p$ is true and 0 otherwise.)
While \tdd instances \mbox{$x=\langle d_\mathrm{issue},c_\mathrm{old}\rangle$}
look the same as SWE-bench instances,
\tdd solutions are functions $\mathit{genTests}$ instead of functions
$\mathit{genCode}$, and it uses an evaluation function
$\mathit{tddScore}$ instead of $\mathit{pass}$.
The evaluation function $\mathit{tddScore}$ uses the hidden golden
new code $\hat{c}_\mathrm{new}$ but it does not use the
hidden golden tests $\hat{y}$.

\subsection{\tdd Evaluation Harness}\label{sec:eval_harness}

\begin{figure}[t]
  \centerline{\includegraphics[width=.95\columnwidth]{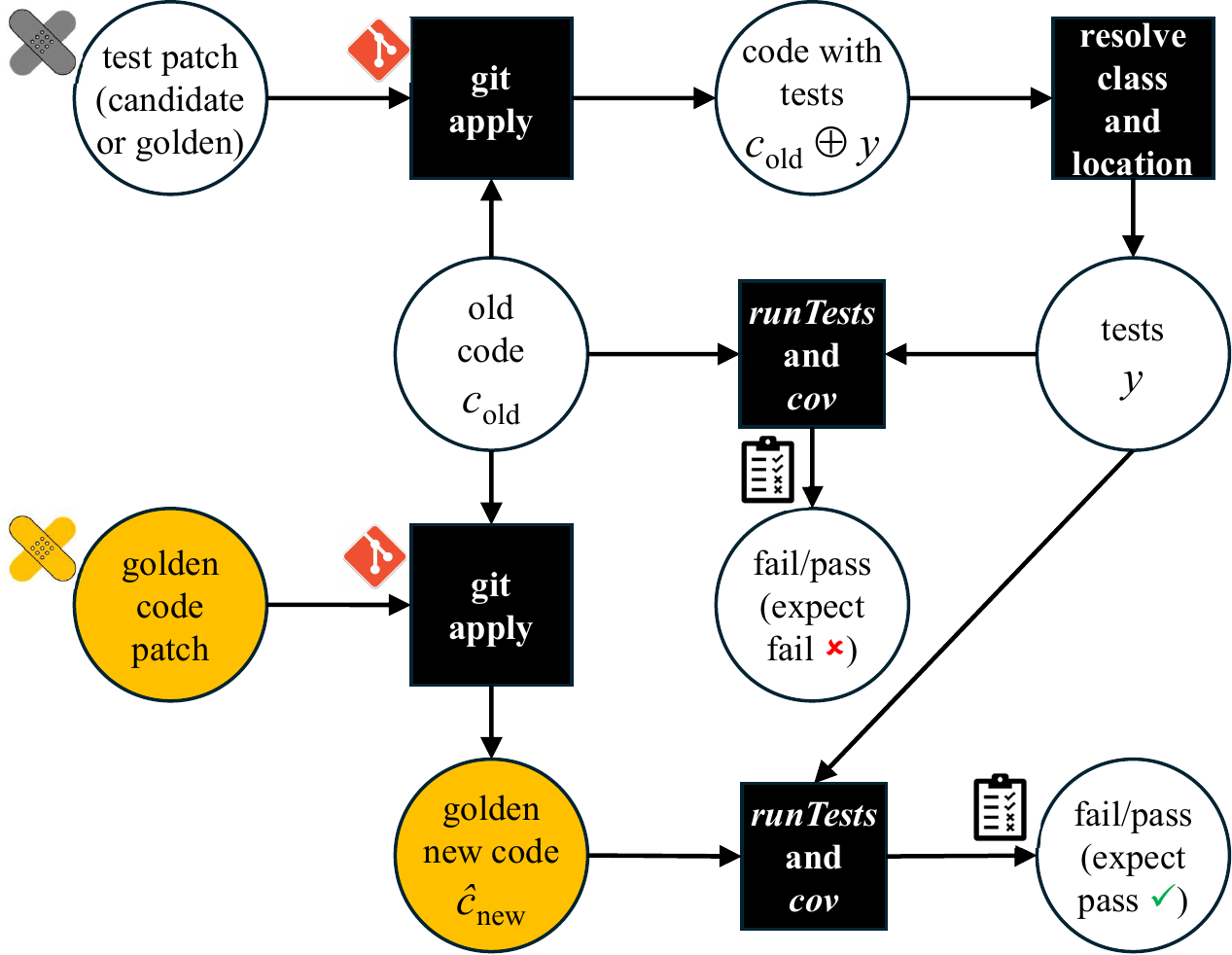}}
  \caption{\label{fig:eval_harness}Evaluation harness for \tdd.}
  \vspace{-15pt}
\end{figure}

Fig.~\ref{fig:eval_harness} shows the harness for evaluating
tests, which typically come from a solution $\mathit{genTests}$
but we can also apply the harness on golden tests from a PR.
The evaluation harness runs in a containerized environment.
Starting at the top left, tests come in the form of a patch, which git
applies on top of the old code $c_\mathrm{old}$.
Next, the harness analyzes the resulting code $c_\mathrm{old}\oplus y$
to resolve the exact list of contributed test functions~$y$.
Once this resolution step is done, the harness can execute the exact
tests $y$ without accidentally running any other tests that happen to
be in the same file but were not part of the test patch.
This yields test results including coverage of the
contributed tests on the old code.
At least one of those results should be a failure for the tests to be
relevant to the issue at hand.

\begin{figure}[t]
    \centering
    \includegraphics[trim=0 0cm 0 0cm, width=\columnwidth]{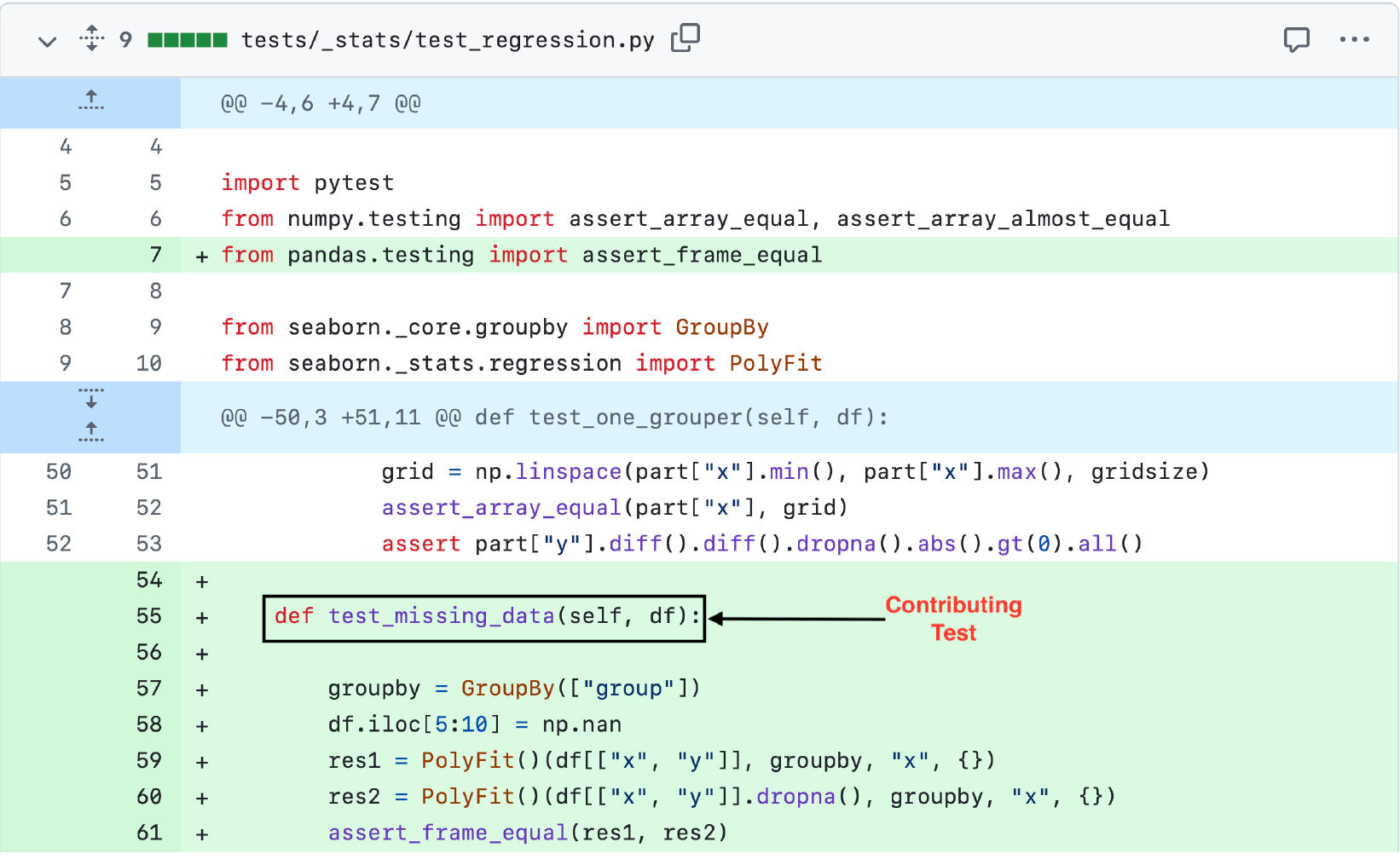}
    \caption{Example test patch with one contributed test. Although the test file name \texttt{\small test\_regression.py} and test name \texttt{\small test\_missing\_data} are available in this text, the class \texttt{\small TestPolyFit} enclosing \texttt{\small test\_missing\_data} is missing. By applying the test patch to the base commit and parsing the file, we retrieve \texttt{\small TestPolyFit}, which is required to run \texttt{\small test\_missing\_data}.}
    \label{fig:test_patch}
    \vspace{-15pt}
\end{figure}

Moving on to the bottom half of Fig.~\ref{fig:eval_harness}, the code changes come
from the golden code patch mined from the same PR, which git applies
to obtain the new code $\hat{c}_\mathrm{new}$.
The harness executes the tests $y$ again, this time on the new code,
to obtain a second set of test results.
This time, all tests should pass, to validate that the issue was
indeed resolved. An example test patch is presented in~\cref{fig:test_patch}.

\subsection{Dataset Filters}\label{sec:tdd_filters}

SWE-bench uses filters to only keep mined instances $x$ for which
$\hat{y}$ contains at least some tests that fail on $c_\mathrm{old}$
and pass on the golden new code $\hat{c}_\mathrm{new}$ from the same
mined PR.
SWE-Bench Verified is a subset of the original instances from
SWE-Bench, consisting of 500 instances vetted by human
annotators~\cite{chowdhury_et_al_2024}.
The annotators filtered out instances where the issue description
$d_\mathrm{issue}$ was underspecified or where the golden tests
$\hat{y}$ were too specific, i.e., would reject some valid
new code~$c_\mathrm{new}$.
They also removed some instances where tests failed due to
environment problems instead of the solution.

In the same spirit, \tdd applies further filters to obtain an even
higher-quality subset of instances.
In a nutshell, the filtering process applies the \tdd evaluation
harness described above to the supposedly golden tests $\hat{y}$
from the original PR.
Specifically, substituting $\hat{y}$ wherever $y$ occurs in
Fig.~\ref{fig:eval_harness} checks whether the PR indeed contributed
tests that went from failing to passing.
We filter out any instance where the contributed tests do not satisfy
that criterion.
As it turns out, while the human annotators of SWE-bench Verified were
diligent, a few instances slipped past their filters, and we drop
those for \tdd.

\cref{fig:tdd-flow} visualizes this filtering process.
Starting from the original 500 instances of SWE-bench verified, we
first drop 7 instances whose environment we could not recreate.
Next, we run the test harness on the golden tests $\hat{y}$.
This filters out 44 additional instances because the tests do not have
the expected fail-to-pass behavior or have zero line coverage on the
golden code patch.
In the end, 449 high-quality instances remain. We summarize key statistics of \tdd in~\cref{tbl:tdd-stat}.

\begin{table*}[t]
\centering
\caption{Different attributes of the \tdd instances.}
\resizebox{.9\textwidth}{!}{%
\renewcommand{\arraystretch}{1.2}
\begin{tabular}{lrrrrrrr}
\toprule
\multicolumn{1}{c}{\multirow{2}{*}{Project}} & \multicolumn{1}{c}{\multirow{2}{*}{\# of Instances}} & \multicolumn{1}{c}{\multirow{2}{*}{\begin{tabular}[c]{@{}c@{}}Fraction of \\ Dataset (in \%)\end{tabular}}} & \multicolumn{1}{c}{\multirow{2}{*}{\# of Files}} & \multicolumn{1}{c}{\multirow{2}{*}{\# of Test Files}} & \multicolumn{2}{c}{\begin{tabular}[c]{@{}c@{}}Average \# of Lines \\ Deleted and Added\end{tabular}} & \multicolumn{1}{c}{\multirow{2}{*}{\begin{tabular}[c]{@{}c@{}}Average Word Count\\ in Issue Description\end{tabular}}} \\ \cmidrule{6-7}
\multicolumn{1}{c}{}                          & \multicolumn{1}{c}{}                                & \multicolumn{1}{c}{}                                                                                        & \multicolumn{1}{c}{}                             & \multicolumn{1}{c}{}                                  & \multicolumn{1}{c}{On Code}                         & \multicolumn{1}{c}{On Tests}                        & \multicolumn{1}{c}{}                                                                                                   \\ \midrule
Astropy                                       & 18                                                  & 4.0                                                                                                        & 1,990                                             & 351                                                   & 11.9                                                & 28.7                                                & 304.5                                                                                                                  \\
Django                                        & 212                                                 & 47.2                                                                                                       & 6,863                                             & 810                                                   & 12.0                                                  & 24.7                                                & 145.6                                                                                                                  \\
Flask                                         & 1                                                   & 0.2                                                                                                        & 275                                              & 27                                                    & 3.0                                                   & 5.0                                                  & 35.0                                                                                                                     \\
Matplotlib                                    & 32                                                  & 7.1                                                                                                        & 4,656                                             & 102                                                   & 9.3                                                 & 20.0                                                  & 260.5                                                                                                                  \\
Pylint                                        & 10                                                  & 2.2                                                                                                       & 3,833                                             & 51                                                    & 24.7                                                & 33.8                                                & 347.1                                                                                                                  \\
Pytest                                        & 16                                                  & 3.6                                                                                                       & 639                                              & 114                                                   & 24.6                                                & 53.5                                                & 250.1                                                                                                                  \\
Requests                                      & 5                                                   & 1.1                                                                                                       & 155                                              & 9                                                     & 3.6                                                 & 6.6                                                 & 85.2                                                                                                                   \\
Scikit-learn                                  & 25                                                  & 5.6                                                                                                        & 1,772                                             & 242                                                   & 11.8                                                & 17.1                                                & 297.6                                                                                                                  \\
Seaborn                                       & 2                                                   & 0.5                                                                                                        & 353                                              & 34                                                    & 13.5                                                & 18.5                                                & 182.5                                                                                                                  \\
Sphinx                                        & 41                                                  & 9.1                                                                                                        & 1,917                                             & 137                                                   & 17.5                                                & 26.1                                                & 186.2                                                                                                                  \\
Sympy                                         & 67                                                  & 14.9                                                                                                       & 2,050                                             & 617                                                   & 12.1                                                & 11.9                                                & 114.2                                                                                                                  \\
Xarray                                        & 20                                                  & 4.5                                                                                                       & 394                                              & 67                                                    & 17.1                                                & 24.3                                                & 301.0                                                                                                                    \\ \midrule
Overall                                       & 449                                                 & 100.0                                                                                                         & 24,897                                            & 2,561                                                  & 13.2                                                & 23.3                                                & 182.0     \\ \bottomrule                               \multicolumn{8}{l} {*File counts are based on the main branches of the project (cloned on October 29, 2024).}
                                                                               
\end{tabular}
}
\label{tbl:tdd-stat}
\vspace{-15pt}
\end{table*}

\subsection{Evaluation Metric}
\label{subsec:eval_metric}

This section defines the evaluation metric $\mathit{tddScore}$ of our
benchmark.
Passing a test does not necessarily mean it is adequate to address the
issue.
Aleithan \etal reported that 31.1\% of the passed patches are
suspicious due to weak test cases in SWE-Bench~\cite{aleithan2024swe}.
To ensure test adequacy or relevance, we also compute the coverage of
the submitted test-patch.
One key difference between SWE-Bench and \tdd is that SWE-Bench runs
an entire test file to evaluate the submitted patch, whereas we only
run the contributing tests $y$ retrieved from the test-patch.
Not running other test cases enables us to precisely track the
coverage of the submitted tests.
If the tests are relevant, they should cover the deleted lines in the
base commit $c_\mathrm{old}$ and the added lines in the commit
$\hat{c}_\mathrm{new}$ where the issue was addressed.
We integrated the Python Coverage package into all 12 repositories and
updated the scripts to allow us to run specific test cases and compute
coverage from them.

The function $\mathit{tddScore}$ evaluates
the quality of tests generated by a solution $\mathit{genTests}$
over a set $X=\{x_0,x_1,\ldots\}$ of instances.
It returns a number between 0 and 100, the higher the better.
It is defined as 100 times the arithmetic mean of the per-instance scores:
\[\mathit{tddScore}(X, \mathit{genTests})
  = \frac{100}{|X|}\sum_{x\in X} \mathit{tddScore}\big(x, \mathit{genTests}(x)\big)
\]

Given a set of tests $y=\mathit{genTests}(x)$ submitted for an
instance, the per-instance score is a product of two factors:
\[\mathit{tddScore}(x, y)
  = \mathit{failToPass}(x, y) \cdot \mathit{adequacy}(x, y)
\]

The first factor is a binary correctness metric, using the indicator
function for the tests $y$ failing on the old code times
the indicator function for the tests $y$ passing on the new code.
While the solution $\mathit{genTests}$ only has access to the 
old \mbox{code $c_\mathrm{old}$}, the evaluation metric also
uses the hidden golden new \mbox{code $\hat{c}_\mathrm{new}$} right after the issue
was fixed.
\[\begin{array}{@{}l@{}}\mathit{failToPass}(x, y) =\\
  \;\;I\big(\mathit{fail} \in \mathit{runTests}(y, c_\mathrm{old})\big)
  \cdot I\big(\mathit{fail} \notin \mathit{runTests}(y, \hat{c}_\mathrm{new})\big)
\end{array}\]

The second factor is the adequacy of the tests, defined as a
fraction between 0 and 1 based on test coverage on
the old and new code:
\[\begin{array}{@{}l@{}}\mathit{adequacy}(x, y) =\\
  \displaystyle\;\;
    \frac{  |\mathit{cov}(y, c_\mathrm{old}) \cap (c_\mathrm{old}\setminus\hat{c}_\mathrm{new})|
          + |\mathit{cov}(y, \hat{c}_\mathrm{new}) \cap (\hat{c}_\mathrm{new}\setminus c_\mathrm{old})|}
         {  |c_\mathrm{old}\setminus\hat{c}_\mathrm{new})| \quad
          + \quad\;\;\, |\hat{c}_\mathrm{new}\setminus c_\mathrm{old}|}
\end{array}\]

Adequacy focuses on just the coverage of lines added and deleted
when going from the old code to the new code, because those are
the most relevant lines to be tested.
In the above, $\mathit{cov}(y,c)$ is the set of lines
covered by running tests $y$ on code~$c$;
$(c_\mathrm{old}\setminus\hat{c}_\mathrm{new})$ is the set
of lines deleted by the PR patch;
and $(\hat{c}_\mathrm{new}\setminus c_\mathrm{old})$ is the set
of lines added by the PR patch.
We had initially considered defining adequacy with two separate
fractions for the deleted vs.\ added lines.
However, that was not only poorly weighted but brittle, because in some cases,
the numerator or denominator of one of the fractions was zero.
Later in the paper, we will give an example for how adequacy
can vary between hand-written tests and LLM-generated tests.


\section{Test Generation from Issue Descriptions}
\label{sec:baselines}

This section first presents a baseline test-generation technique, which is a zero-shot approach for generating \emph{test files} from issue descriptions. Then, it presents \atdd, which implements a few-shot approach for generating \emph{test functions} from issue descriptions.

\begin{figure}[!t]
    \centering
    \includegraphics[width=\columnwidth]{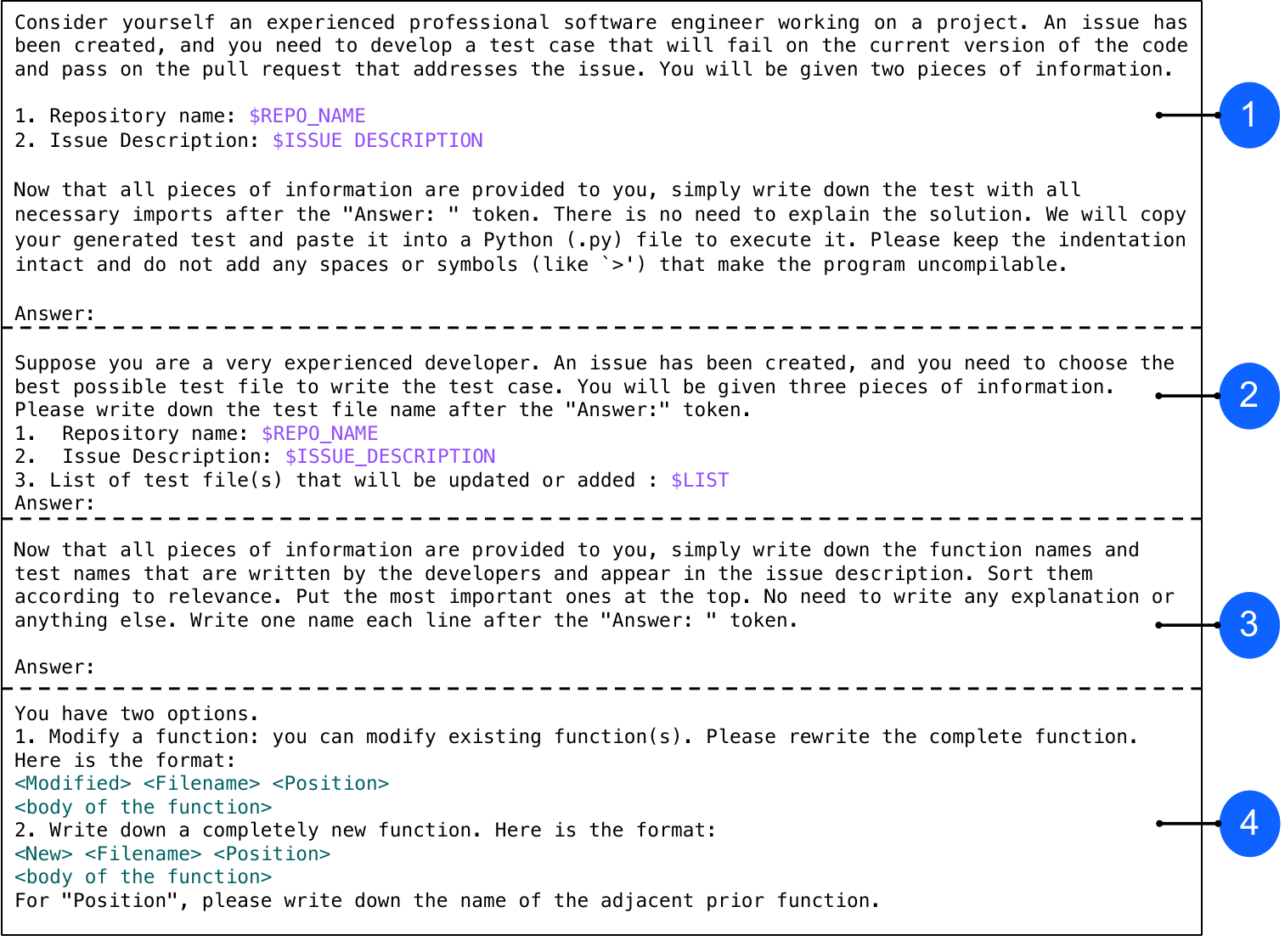}
    \caption{Different prompts used in our approach: \circled{1} shows the prompt for the baseline zero-shot test file generation; \circled{2}--\circled{4} show the prompts used for the sequence of LLM calls (for selecting a test file, identifying issue-related test functions, and generating test functions) in \atdd.}
    \label{fig:prompt}
    \vspace{-14pt}
\end{figure}

\subsection{Baseline: Zero-shot Test File Generation}
Recent instruction-tuned LLMs excel at following instructions~\cite{peng2023instruction,zhang2023instruction}. We start with a simple zero-shot approach to generate a fail-to-pass test given the repository name and the issue description (see detailed prompt in~\cref{fig:prompt}-\circled{1}). Given the prompt, the model generates a complete test file with all necessary imports to make it compilable. While in real scenarios, test files usually have multiple test cases, this baseline usually generates only a single test per instance. The generated test file (we call it \texttt{\small test\_tdd.py}) needs to be placed in the right directory because otherwise, the imports in the test may not work. Fortunately, all Python projects in \tdd have at least one directory called \texttt{\small tests}; some projects have multiple such directories. So we follow the simple approach of searching for the \texttt{\small tests} directory and placing \texttt{\small test\_tdd.py} in that directory. 

\subsection{\atdd: Few-shot Test Function Generation}
\label{method:autotdd}

One of the caveats of the baseline approach is that it does not support the typical practice of developers while addressing an issue---that they do not write a new test file but instead update an existing test file, by adding tests to it or modifying tests in it (the SWE-Bench test patches strongly support this hypothesis). Working on an existing file helps the developers with necessary imports and all the dependencies, thus simplifying their task. To support in-file test-function insertion, we need a better approach that localizes the most suitable test file and the position in the file at which to insert the generated tests.
\atdd uses a pipeline of three LLM calls.
The first LLM call uses a zero-shot prompt to select a test file.
The second LLM call uses a zero-shot prompt to guess issue-related
functions (discussed below).
The third LLM call uses a few-shot prompt to generate a new test
function and pick a position for inserting it in the test file.

\begin{figure}[!t]
    \centering
    \includegraphics[width=\columnwidth]{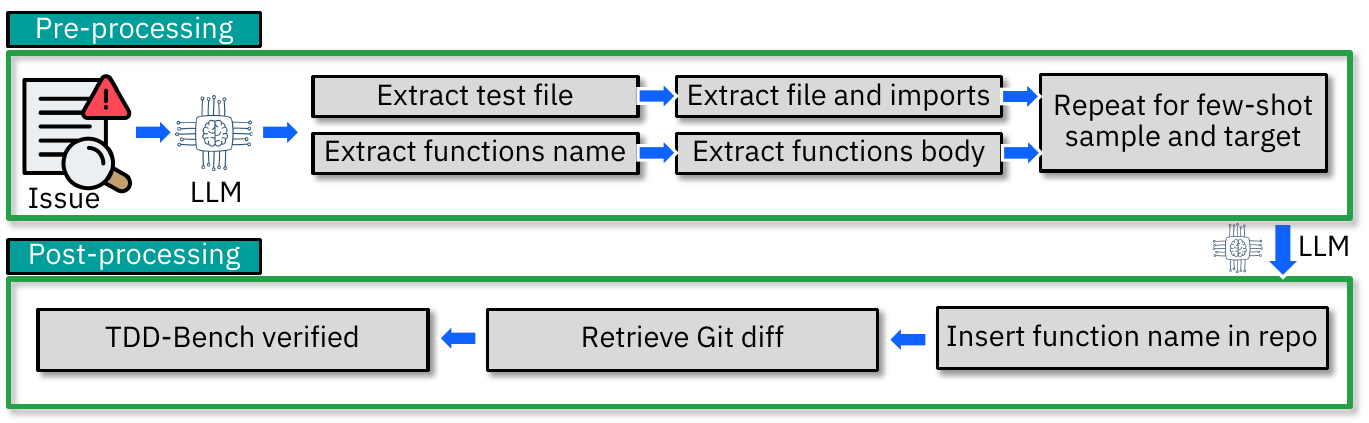}
    \caption{Overall architecture of \atdd.}
    \label{fig:few-shot}
    \vspace{-15pt}
\end{figure}

Few-shot learning is popular in natural language processing (NLP)~\cite{brown2020language}. Few-shot learning and its variants are also widely applied in developing techniques for various software-engineering tasks, such as unit test generation~\cite{bareiss2022code}, code summarization~\cite{ahmed2022few, ahmed2024automatic}, program repair~\cite{nashid2023retrieval}, and program synthesis~\cite{misu2024towards}. In few-shot learning, we do not update the model's parameters; instead, we make the model perform a certain task by providing multiple examples in the context. The idea is to present the model with a certain number of input-output pairs and ask the model to provide the solution for the last instance.
Research has shown that few-shot learning is generally superior to zero-shot learning for most tasks. Besides, in few-shot learning, it is easy to instruct the model to generate the output in a specific format, which is important for post-processing the results. 

\subsubsection{Selecting a test file}\label{sec:test_file_selection}
Selecting the correct test file can be crucial for improving test generation performance.
The third LLM call of \atdd is intended to generate only the test function, without knowing the necessary imports and dependencies, so we should not expect the model to handle those aspects.
If the first LLM call can correctly select the test file, the third call can thus benefit from the imports.
Additionally, as part of the approach, we provide the model with a file name, its classes, and methods to determine where to insert into or modify the test file. Because we do not know (or cannot infer from the benchmark) the correct test file, we should apply some methods to find it.
To do so, we first navigate to the base commit $c_\textrm{old}$ where the issue was created and collect all files that contain at least one test.

In an initial attempt, we took the list of test files and tried to find any name present in the issue description~$d_\mathrm{issue}$.
This conclusively found a test file for only 3 instances~($<$1\%). However, even when the test file name does not appear directly, there are enough hints in the issue description that could be leveraged to guess the correct test file. For example, instance astropy\_\_astropy-12907\footnote{\url{https://github.com/astropy/astropy/issues/12906}} does not contain the test file name \texttt{\small astropy/modeling/tests/test\_compound.py} directly, but it contains multiple mentions of terms like ``model'' and ``compound'' that an LLM can leverage to guess the correct test file.
We tried a simple zero-shot approach, where we presented the model with the issue description $d_\mathrm{issue}$ and names of test files in $c_\mathrm{old}$ and asked the model to choose a suitable test file (see detailed prompt in \mbox{\cref{fig:prompt}-\circled{2}}). \cref{tbl:bm25} shows that LLMs are good at identifying test files from the issue description (56\%--62\% accuracy, depending on the model) and perform much better than a traditional frequency-based retrieval algorithm like BM25 (only 15\% accuracy). 

\begin{table}[t]
\centering
\caption{BM25-based and model-based test file retrieval.}
\resizebox{.70\columnwidth}{!}{%
\renewcommand{\arraystretch}{1.2}
\begin{tabular}{lr@{\hspace*{12mm}}r@{\hspace*{5mm}}}
\toprule
\multicolumn{1}{c}{Approach} & \multicolumn{1}{c}{\begin{tabular}[c]{@{}c@{}}\# of Correct Retrieval \\ (Out of 449)\end{tabular}} & \multicolumn{1}{c}{\begin{tabular}[c]{@{}c@{}}Accuracy \\ (in \%)\end{tabular}} \\ \midrule
BM25                         & 69                                                                                                  & 15.4                                                                            \\ \midrule
Llama-3.1                    & 253                                                                                                 & 56.3                                                                            \\
Mistral-Large                & 278                                                                                                 & 61.9                                                                            \\
GPT-4o                       & 276                                                                                                 & 61.5  \\ \bottomrule                                                                          
\end{tabular}
}
\label{tbl:bm25}
\vspace{-.5cm}
\end{table}




\subsubsection{Guessing issue-related functions}\label{sec:issue_related_functions}
Extracting issue-related functions could be helpful for the LLM in generating the test function.
The issue description $d_\mathrm{issue}$ often contains names of functions.
However, there are two challenges: (1)~some functions are library functions that do not have any project-specific information, and (2)~it is hard to parse the function names from the issue description because it is written in natural language.
Hence, we perform a second LLM call to parse these function names.
This step uses a zero-shot prompt (see detailed prompt in \mbox{\cref{fig:prompt}-\circled{3}}), and the LLM returns a list of functions.
Then, we look into the repository for those functions and, if we find them, we simply add them into the context as relevant functions.

\subsubsection{Generating and inserting a test function}

This is the third and most important LLM call in \atdd.
The prompt starts with a guideline that instructs the model about context information provided and the expected output format. After the guideline, the prompt contains three context-solution pairs as few-shot samples. 
Each context provides four pieces of information: (1)~repository name, (2)~issue description $d_\mathrm{issue}$, (3)~issue-related functions from \cref{sec:issue_related_functions}, and (4)~a skeleton of the test file from \cref{sec:test_file_selection} containing the test class names, test method declarations, and imports.
It is important to know the name of the test file and be aware of the adjacent function to locate where to insert the test.

We instruct the model to follow a specific output format in the guideline and also demonstrate some samples in the context. In terms of tests, we can have two distinct scenarios: (1)~the LLM modifies an existing test, and (2)~the LLM writes a completely new test. To handle these cases, the prompt instructs the LLM to start its response with ``Modified'' or ``New'' to indicate the type of test case (see detailed prompt in~\cref{fig:prompt}-\circled{4}). The LLM also generates the test file name and position of the generated test. For modification, we try to find the function and replace the existing one with the new one. If the function cannot be found, we add the new function at the end of the file. We also adjust the indentation by looking at the original function or the prior function if we fail to find the function. For a newly written function, we expect the model to generate the name of the adjacent existing function. We find that function in the file and insert the newly written function right after that. For a newly written function, the model is allowed to write ``first'' to indicate that the function can be written at the beginning of the file, right before the existing first functions. Like modification, in this case, we also repair the indentation and have a fallback plan with inserting the function at the end if anything goes wrong or the model hallucinates any name. Note that we could generate a patch from the model. However, from our initial experiment, we found that the model hallucinates the line number of the patch and could not be applied to the original code for evaluation. Therefore, instead of asking the LLM to generate a patch, we ask it to generate a function, which we insert into the repository after the LLM call. Then, we just use the \texttt{\small git diff} tool to obtain a patch that can be used for evaluation with \tdd. \cref{fig:few-shot} presents the overall architecture of \atdd.


\section{Evaluation Methodology}
\label{sec:method}

We designed the evaluation to answer the following research questions:

\begin{description}
\item[\textbf{RQ1:}] How does \atdd perform in generating fail-to-pass tests on \tdd instances?
\item[\textbf{RQ2:}] How do different components of \atdd contribute to its performance?
\item[\textbf{RQ3:}] How does \atdd perform in the ``write first, test later'' setup?
\item[\textbf{RQ4:}] What is the coverage adequacy of developer-written tests and \atdd-generated tests?
\end{description}







The rest of this section briefly discusses the dataset, models, and methodology used to answer the research questions.  Section~\ref{sec:result} presents the results of this evaluation.

\subsection{Dataset and Models}
The evaluation dataset consists of the 449 instances of \tdd discussed in \cref{sec:tddbench}.
These instances belong to 12 popular Python repositories. We also collected three samples from the SWE-Bench Dev~\cite{jimenezswe} split to use as few-shot samples, which belong to 3 different repositories disjoint from the 12 test-split repositories. 

We selected three models for our experiments: Llama-3.1, Mistral-large, and GPT-4o.

\vspace{.2cm}
\noindent{\emph{Llama-3.1:}} Llama 3.1 is a multilingual instruction-tuned LLM. It is an auto-regressive language model that uses an optimized transformer architecture. The models are tuned by supervised fine-tuning (SFT) and reinforcement learning with human feedback (RLHF) to align with human preferences. It outperforms many of the available open-source and closed chat models on several benchmarks. We used the 70 billion parameter model with 128K context window for our experiments.

\vspace{.2cm}

\noindent{\emph{Mistral-large:}} Mistral-large is an instruction-tuned LLM with 123 billion parameters. It has state-of-the-art reasoning, knowledge, and coding capabilities. By design, the model is multilingual, proficient in coding, and possesses agentic capabilities. Additionally, it features a large context window of 128K.

\vspace{.2cm}

\noindent{\emph{GPT-4o:}} GPT-4o is a recent frontier model by OpenAI. Unlike Llama-3.1 and Mistral-Large, which are open-source models, GPT-4o is a closed-source model; the model size is not known publicly.
GPT-4o is multimodal and can accept text or image inputs and output text. It is also very good at coding with a context window of 128K tokens. We use the temperature value~0 and maximum output of 4096 tokens to keep the configuration similar to the other models.

\subsection{Methodology for RQ1 (effectiveness of \atdd)} 
\label{sec:rq1-method}

In the first research question, we investigate the effectiveness of the baseline approach and \atdd described in~\cref{sec:baselines}. 
Note that LLMs (especially the smaller models) sometimes generate unparseable code. These models also have an inclination to generate natural language descriptions with the test, which is sometimes unavoidable even with explicit instructions. Also, if the models generate solutions with wrong format, we cannot process those samples. We report the number of samples we lost for various reasons. After that, we compute the number of instances for which each model generates failing test(s). Some tests may pass even on the prior version $c_\mathrm{old}$ of the code; such tests are irrelevant for us. Eventually, we report the number of failing and passing instances in PR commits $\hat{c}_\mathrm{new}$ along with the fail-to-pass  rate \mbox{$\displaystyle\frac{1}{|X|}\sum_{x\in X}\mathit{failToPass}(x, \mathit{genTests}(x))$} and the final score $\mathit{tddScore}(X, \mathit{genTests})$, which also factors in test adequacy.

We also compare our approach with the approaches proposed by M{\"u}ndler \etal~\cite{mundler2024code}.
Their strongest approach, \mbox{SWE-Agent+}, is the prior state of the art for our task.
However, their paper evaluates on a different set of 276 instances, which
is a subset of SWE-Bench Lite, and not verified.
Therefore, for this comparison, we use their 276 instances instead of our 449 instances.
Note that \atdd runs only the contributed tests, whereas M{\"u}ndler \etal run the complete test file. However, we believe the fail-to-pass metric is still comparable.
The sole purpose of this experiment is to compare with the state-of-the-art existing technique.

\subsection{Methdology for RQ2 (ablation study for \atdd)} 
As described in~\cref{method:autotdd}, \atdd augments the context with test file structure and imports, as well as relevant functions. In this research question, we perform an ablation by removing each component individually to show how the performance degrades. We start with the relevant functions, followed by the test file imports. It is essential to have at least the test file structure to make \atdd work. To show the importance of LLM-based file selection, we replace the LLM-selected file with the file selected by BM25 retrieval (\cref{tbl:bm25}).

\subsection{Methodology for RQ3 (``write first, test later'')}
\label{method-rq3}

While TDD has been well-known to the community for several years now~\cite{beck_2002}, it has both pros and cons. Adopting TDD can be challenging for developers with less experience in using unit testing or writing modular code. Thus, the traditional approach of  ``write first, test later’’ is still popular. Although our benchmark is primarily aimed at TDD, it can be easily adapted for this approach. Writing tests is a challenging task either way, and it would be beneficial to write tests for both approaches.
To accommodate the ``write first, test later’’ approach, instead of providing the solution with only the base commit $c_\mathrm{old}$, we can also provide it with the commit $\hat{c}_\mathit{new}$ that addresses the issue.
To ensure the validity of the submitted golden patch, the test still needs to fail on the base commit and pass on the later commit. For our baseline, we made a small change by adding the code patch (computed via \texttt{\small git diff}) that addresses the issue to the three pieces of information used in the prompt. This way, we can convey the changes made to address the issue. Alternative approaches could be explored to convey this information, but this is beyond the scope of this paper. In short, we repeat the same methodology we followed in RQ1, but we include the code patch in the prompt.

\subsection{Methodology for RQ4 (adequacy and hand-written tests)}

\begin{figure}[t]
    \centering
    \begin{subfigure}{\columnwidth}
        \centering
        \includegraphics[width=0.7\columnwidth]{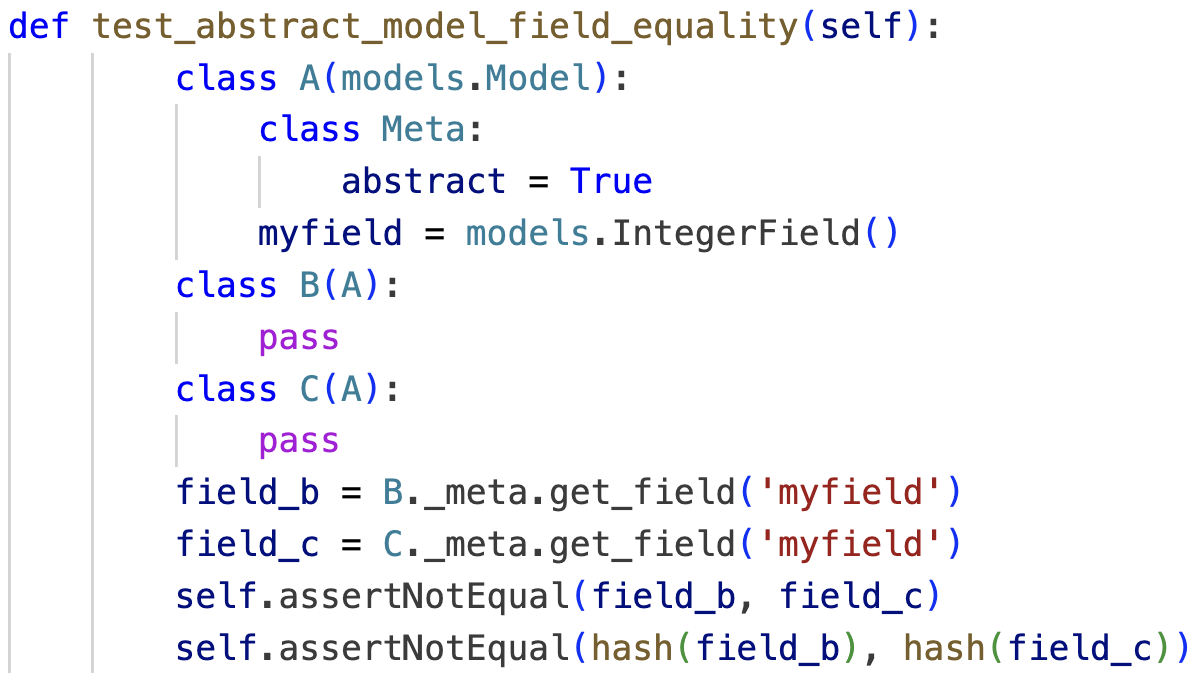}
        \caption{Test generated by GPT-4o}
        \label{fig:sub1}
    \end{subfigure}
    ~\\
    \begin{subfigure}{\columnwidth}
        \centering
        \includegraphics[width=\columnwidth]{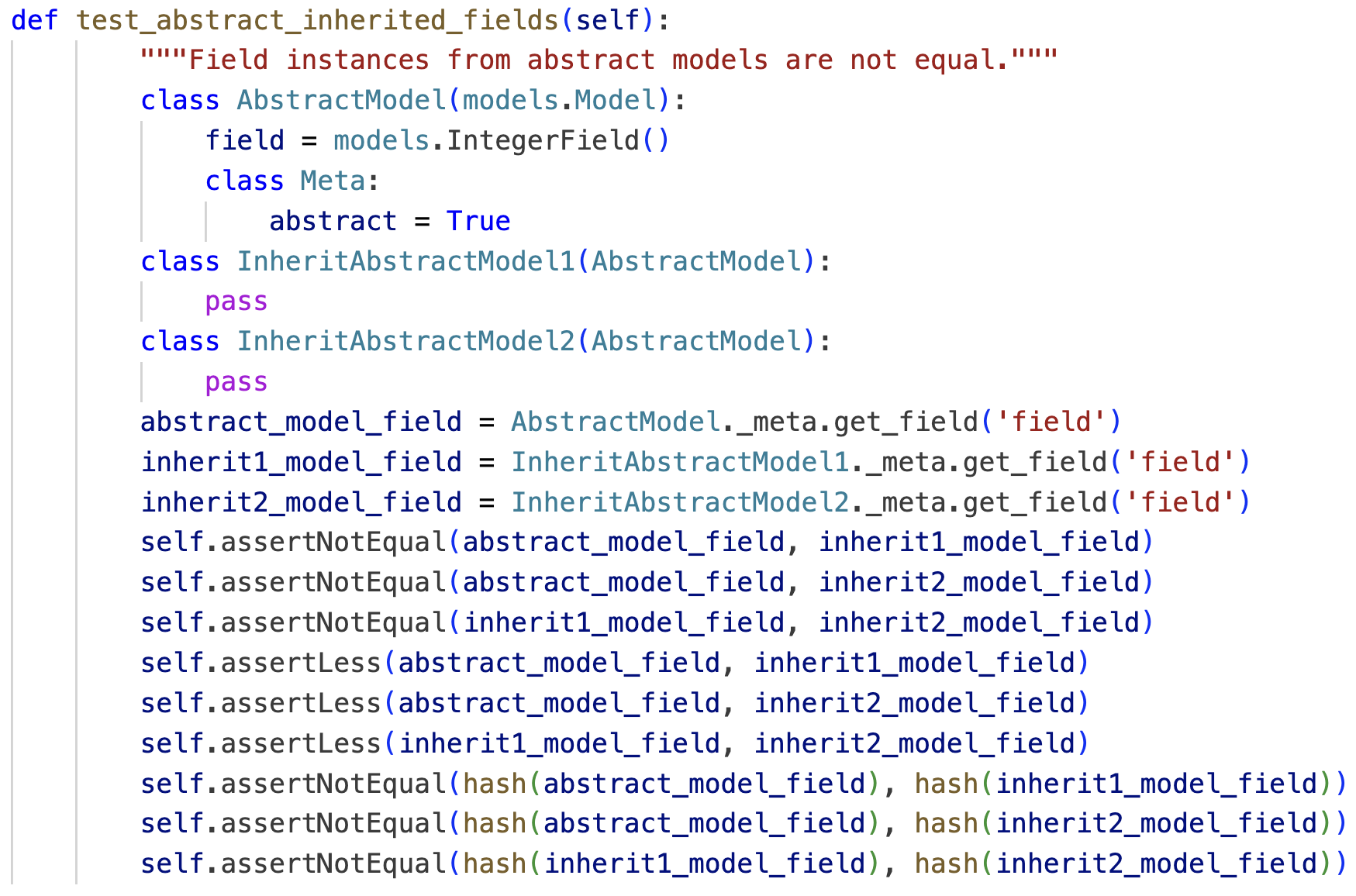}
        \caption{Developer-written test}
        \label{fig:sub2}
    \end{subfigure}
    \caption{Model-generated and developer-written fail-to-pass test for addressing the same issue (django\_\_django-13401).}
    \label{fig:coverage}
    \vspace{-15pt}
\end{figure}

Although tests are useful for finding bugs, they can be inadequate. To measure adequacy, we use code coverage, a widely used metric.
\cref{subsec:eval_metric} explains how our new benchmark, \tdd, handles coverage in its evaluation metric and evaluation harness.
\cref{fig:coverage} shows two test samples: one written by a developer and the other generated by GPT-4o. Both tests transition from fail to pass after applying the golden patch. But, are they equally adequate or good? The model-generated test is not necessarily as good as the human-written one. The original issue (django\_\_django-13401\footnote{\url{https://github.com/django/django/pull/13401}}) was that fields from abstract models are considered equal across different models, which can lead to unexpected behavior when using sets or other data structures that rely on equality comparisons.
The GPT-4o-generated test only asserts that the fields are not equal, whereas the developer-written test also asserts that the fields are less than each other and that their hash values are not equal. The developer-written test is thus more comprehensive as it also checks the comparison of fields and their hash values. 

The coverage metric is a good indicator of such quality distinctions. For example, the coverage for these two tests is 0.71 and 0.96, respectively. This research question investigates the adequacy of human-written tests~$\hat{y}$ and model-generated tests~$y$. We are interested in knowing whether they have similar and sufficient adequacy. To do so, we collect the coverage information of model-generated fail-to-pass tests and compare them with the human-written tests. To further break down the results, we follow two steps. First, we measure the coverage of model-generated fail-to-pass tests only and compare them with developer-written tests. Second, we compare coverage for other tests separately to see the coverage difference between fail-to-pass tests and other tests~(i.e., fail-to-fail, pass-to-fail, or pass-to-pass).

\section{Evaluation Results}
\label{sec:result}
This section discusses the results and our findings on the four research questions.

\subsection{RQ1: Performance of Our Baseline and \atdd}
\begin{table*}[t]
\centering
\caption{Performance of the baseline technique and \atdd on \tdd.}
\resizebox{.75\textwidth}{!}{%
\renewcommand{\arraystretch}{1.2}
\begin{tabular}{llrrrrrr}
\toprule
\multicolumn{1}{c}{\multirow{2}{*}{Model}} & \multicolumn{1}{c}{\multirow{2}{*}{Technique}} & \multicolumn{1}{c}{\multirow{2}{*}{\begin{tabular}[c]{@{}c@{}}Syntactic Errors \\ or Formatting Issues\end{tabular}}} & \multicolumn{1}{c}{On $c_\mathrm{old}$} & \multicolumn{2}{c}{On $\hat{c}_\mathrm{new}$} & \multicolumn{1}{c}{\multirow{2}{*}{\begin{tabular}[c]{@{}c@{}}\# of Fail-to-Pass\\ Tests in (\%)\end{tabular}}} & \multicolumn{1}{c}{\multirow{2}{*}{$\mathit{tddScore}$}} \\ \cline{5-6}
\multicolumn{1}{c}{}                       & \multicolumn{1}{c}{}                        & \multicolumn{1}{c}{}                                                                                                  & \multicolumn{1}{c}{Fail}                                                      & \multicolumn{1}{l}{Fail}             & \multicolumn{1}{l}{Pass}             & \multicolumn{1}{c}{}                                                                                                & \multicolumn{1}{c}{}                             \\ \midrule
\multirow{2}{*}{Llama-3.1}                 & Zero-shot                                   & 69                                                                                                                    & 243                                                                           & 192                                  & 51                                   & 11.4                                                                                                                & 10.3                                             \\
                                           & Auto-TDD                                     & 75                                                                                                                    & 331                                                                           & 294                                  & 37                                   & 8.2                                                                                                                 & 7.8                                              \\ \midrule
\multirow{2}{*}{Mistral-Large}             & Zero-shot                                   & 52                                                                                                                    & 292                                                                           & 235                                  & 57                                   & 12.7                                                                                                                & 11.8                                             \\
                                           & Auto-TDD                                     & 11                                                                                                                    & 395                                                                           & 318                                  & 85                                   & 18.9                                                                                                                & 18.3                                             \\ \midrule
\multirow{2}{*}{GPT-4o}                    & Zero-shot                                   & 52                                                                                                                    & 273                                                                           & 189                                  & 84                                   & 18.7                                                                                                                & 17.2                                             \\
                                           & Auto-TDD                                     & 15                                                                                                                    & 392                                                                           & 286                                  & 106                                  & 23.6                                                                                                                & 22.6      \\ \bottomrule                                       
\end{tabular}
}
\label{tbl:tdd-performance}
\vspace{-7pt}
\end{table*}

For zero-shot test file generation, we have three models, and we used the exact same prompts for all of them. We were able to generate fail-to-pass tests for 51, 57, and 84 instances for the Llama-3.1, Mistral-large, and GPT-4o models, respectively (see~\cref{tbl:tdd-performance}). The values for $\mathit{tddScore}$ are quite close to the percentages of fail-to-pass instances (e.g., 18.7\% vs.\ 17.2\% for GPT-4o). 
As discussed earlier, we consider coverage in our final score. Because a few fail-to-pass tests do not have perfect coverage, it is slightly lower than the percentage of instances. We discuss more about coverage in \cref{sec:result-rq2}.

In \atdd, we also used the structure of a test file to determine the location of the test case. For file localization, we use the model under consideration to select a file to write the test, and the models are quite good at this task, achieving 56\%--62\% Top-1 accuracy (see~\cref{tbl:bm25}). 
We also ask the model to extract relevant function names from the issue description and search for those functions in the repository. If we find the functions, we simply add them to the context. 
Three samples are taken from the SWE-Bench Dev split to be used as few-shot examples. These are from three completely different repositories, and there is little chance that the model will be biased by these three samples. For two models, GPT-4o and Mistral-large, \atdd achieved 20--22 more fail-to-pass tests compared to our zero-shot approach. However, for Llama-3.1, \atdd generates 14 fewer fail-to-pass tests than the baseline approach. There is a possibility that Llama-3.1 is being misled by the context provided for the few-shot samples. Note that the goal of few-shot samples is not to add related information here, but rather to force the model to follow a specific format for post-processing.



\vspace{.2cm}

\noindent\emph{Distribution of generated tests:} We achieved 7.8\%--22.6\% final scores using different models (Column~8 of~\cref{tbl:tdd-performance}). We analyzed the distribution of passing and failing tests to see how many samples actually go through our whole pipeline without getting dropped because of syntactic errors. Since our full process depends on parsing and indentation of the program (it's Python!), we can lose a few tests in intermediate steps for not fully aligning with our expected format.
Column~3 of~\cref{tbl:tdd-performance} shows that we lost 52--69 samples for the zero-shot file generation baseline that way. Weaker models sometimes generate code that is not parseable or do not produce solutions in the expected format. However, GPT-4o and Mistral-large are relatively good at generating parseable and well-formatted tests, and we lost fewer samples compared to Llama-3.1. 
We also observe how the model-generated tests perform before and after the insertion of the golden code patch $\hat{c}_\mathrm{new}$ (that addresses the issue). The results show that a large number of tests fail on $c_\mathrm{old}$ (243--395), but the number of tests passing after insertion is low (37--106). Note that failure on $c_\mathrm{old}$ is necessary for a test to be relevant for the corresponding issue, and GPT-4o and Mistral generate a lot of failing tests, while generating the maximum number of tests that pass on $\hat{c}_\mathrm{new}$.

\begin{table}[t]
\centering
\caption{Comparing with approaches proposed by M{\"u}ndler \etal~\cite{mundler2024code} on their 276 instances (not \tdd).}
\resizebox{.70\columnwidth}{!}{%
\renewcommand{\arraystretch}{1.2}
\begin{tabular}{lrr}
\toprule
\multicolumn{1}{c}{Approach} & \multicolumn{1}{c}{\begin{tabular}[c]{@{}c@{}}\# of Fail-to-pass \\ Tests\end{tabular}} & \multicolumn{1}{c}{in (\%)} \\ \midrule
ZeroShot                     & 16                                                                                          & 5.8                        \\
ZeroShotPlus*                & 28                                                                                         & 10.1                        \\
LIBRO*~\cite{kang2023large}                       & 42                                                                                         & 15.2                         \\
AutoCodeRover~\cite{zhang2024autocoderover}                & 25                                                                                         & 9.1                         \\
SWE-Agent                    & 46                                                                                         & 16.7                         \\
SWE-Agent+                   & 53                                                                                         & 19.2                       \\ \midrule
\textbf{\atdd}                      & \textbf{60}                                                                                         & \textbf{21.7}  \\ \bottomrule
\multicolumn{3}{l}{* follows ``write first, test later'' approach} 
\end{tabular}
}
\label{tbl:swt}
\vspace{-10pt}
\end{table}


\vspace{.2cm}

\noindent\emph{Comparing with Approaches Proposed by M{\"u}ndler \etal:}
M{\"u}ndler \etal~\cite{mundler2024code} proposed a set of approaches for generating fail-to-pass tests. We ran \atdd on their dataset to study how the approaches compare. They also have zero-shot approaches, which differ from our zero-shot baseline. Instead of generating a complete function, all of their approaches (including zero-shot ones) instruct the model to generate a specific form of ``diff''. Two of their approaches use a golden patch in the prompt, which resembles our ``write first, test later'' setting~(see \cref{method-DDT}). M{\"u}ndler \etal's SWE-agent and SWE-agent+ approaches are derived from SWE-Agent, which was originally designed for generating golden code patches~\cite{sweagent2}. \cref{tbl:swt} shows the results. \atdd performs better than their best-performing approach, generating 60 (21.7\%) fail-to-pass tests compared to 53 (19.2\%) fail-to-pass tests generated by SWE-agent+. Some of this gain comes from our simpler output format and some from our carefully crafted neuro-symbolic pipeline.

\findings{1}{\atdd is able to generate fail-to-pass tests with $\mathit{tddScore}$ between 7.8\% and 22.6\%, depending on the model. \atdd improved performance for GPT-4o and Mistral-large over the baseline, but the performance decreases for Llama-3.1.}

\begin{table}[t]
\centering
\caption{Contribution of each component of \atdd.}
\resizebox{\columnwidth}{!}{%
\renewcommand{\arraystretch}{1.2}
\begin{tabular}{llrrr}
\toprule
\multicolumn{1}{c}{Model} & \multicolumn{1}{c}{Technique} & \multicolumn{1}{@{\hspace*{-3mm}}c}{\begin{tabular}[c]{@{}c@{}}\# of Fail-to-Pass \\ Tests\end{tabular}} & \multicolumn{1}{c}{\begin{tabular}[c]{@{}c@{}}\# of Fail-to-Pass \\ Tests in (\%)\end{tabular}} & \multicolumn{1}{c}{\begin{tabular}[c]{@{}c@{}}Change\\  in (\%)\end{tabular}} \\ \midrule
\multirow{4}{*}{Mistral}  & Auto-TDD                    & 85                                                                                         & 18.9                                                                                               & NA                                                                            \\
                          & Auto-TDD - related functions          & 77                                                                                         & 17.1                                                                                               & -1.8                                                                          \\
                          & Auto-TDD -  import          & 68                                                                                         & 15.1                                                                                               & -3.8                                                                          \\
                          & Auto-TDD - file detection*  & 44                                                                                         & 9.8                                                                                                & -9.1                                                                           \\ \midrule
\multirow{4}{*}{GPT-4o}   & Auto-TDD                    & 106                                                                                        & 23.6                                                                                               & NA                                                                            \\
                          & Auto-TDD - related functions          & 100                                                                                        & 22.3                                                                                               & -1.3                                                                          \\
                          & Auto-TDD -  import          & 96                                                                                         & 21.4                                                                                               & -2.2                                                                          \\
                          & Auto-TDD - file detection*  & 68                                                                                         & 15.1                                                                                               & -8.5     \\ \bottomrule

\multicolumn{5}{l}{* We use BM25 to select the file instead of relying on the model's choice.}                              
\end{tabular}

}
\label{tbl:ablation}
\vspace{-15pt}
\end{table}

\subsection{RQ2: Ablation of \atdd}

\cref{tbl:ablation} shows that each component contributes to the overall performance of \atdd. If we remove the relevant functions, both Mistral-Large and GPT-4o performance go down by 1.78\% and 1.34\% respectively. The import statements are also important and removing them results in 2.23\%--3.79\% performance degradation. However, the most significant component is the right file selection. When we replace the model-selected file with the BM25-retrieved file, the performance drops by 8.47\%--9.13\%. Because the performance of \atdd is worse than the zero-shot baseline with Llama-3.1, we exclude Llama-3.1 from this experiment. 

\findings{2}{Each component of \atdd contributes to its performance. However, LLM-based test file selection plays the most significant role.}

\begin{table}[t]
\centering
\caption{Performance of the zero-shot baseline and of \atdd in the ``write first, test later'' setting.}
\resizebox{\columnwidth}{!}{%
\renewcommand{\arraystretch}{1.2}
\begin{tabular}{llrrr}
\toprule
\multicolumn{1}{c}{\multirow{2}{*}{Model}} & \multicolumn{1}{c}{\multirow{2}{*}{Technique}} & \multicolumn{1}{c}{\multirow{2}{*}{\begin{tabular}[c]{@{}c@{}}\# of Fail-to-Pass \\ Tests\end{tabular}}} & \multicolumn{1}{c}{\multirow{2}{*}{\begin{tabular}[c]{@{}c@{}}\# of Fail-to-Pass \\ Tests in (\%)\end{tabular}}} & \multicolumn{1}{c}{\multirow{2}{*}{\begin{tabular}[c]{@{}c@{}}Final \\ Score\end{tabular}}} \\ 
\multicolumn{1}{c}{}                       & \multicolumn{1}{c}{}                        & \multicolumn{1}{c}{}                                                                                        & \multicolumn{1}{c}{}                                                                                                & \multicolumn{1}{c}{}                                                                        \\ \midrule
\multirow{2}{*}{Llama-3.1}                 & Zero-shot                                   & 64                                                                                                          & 14.3                                                                                                               & 13.3                                                                                       \\
                                           & AutoTDD                                     & 37                                                                                                          & 8.2                                                                                                                & 7.8                                                                                        \\\midrule
\multirow{2}{*}{Mistral-Large}             & Zero-shot                                   & 79                                                                                                          & 17.6                                                                                                               & 16.3                                                                                       \\
                                           & AutoTDD                                     & 97                                                                                                          & 21.6                                                                                                                & 20.8                                                                                       \\\midrule
\multirow{2}{*}{GPT-4o}                    & Zero-shot                                   & 99                                                                                                          & 22.1                                                                                                               & 20.4                                                                                       \\
                                           & AutoTDD                                     & 109                                                                                                         & 24.3                                                                                                               & 23.6    \\ \bottomrule                                                                                   
\end{tabular}
}
\label{tbl:ddt-performance}
\end{table}

\subsection{RQ3: TDD-Bench for ``Write First, Test Later''}
\label{method-DDT}

As mentioned in~\cref{method-rq3}, in the ``write first, test later'' approach, we include the code patch computed using \texttt{\small git diff} in the prompt, presenting the actual code change (not tests) to our techniques and observe how they performs on \tdd. The key takeaway here is that, although \tdd is primarily developed for test-driven development, it can also be applied in other settings. We have slightly better performance in this setup as it provides more context to the model; e.g., for GPT-4o, \atdd generated 109 fail-to-pass tests in this setting compared with 106 without the code patch. One of the interesting observations is that, even in this setup, Llama-3.1's performance goes down with \atdd, whereas we have seen improved performance with both Mistral-Large and GPT-4o, exactly as we have seen with the TDD setup (\cref{tbl:tdd-performance}).
As expected, the final $\mathit{tddScore}$ is also slightly on the higher side: 23.6\% with GPT-4o with \atdd compared to 22.6\% in the TDD setting.
Although we have seen improved performance in the ``write first, test later'' approach, the difference is not very surprising. Note that, in the current evaluation, we used the \texttt{\small git diff} patch format to present the changed code to the model. Alternatively, this information could be presented in a more natural way, such as including the complete function or incorporating a natural language description of the change. However, this paper is primarily focused on the TDD approach, so we leave using different presentation schemes for future research.

\findings{3}{TDD-Bench is also applicable to the ``write first, test later'' approach, and our baselines do slightly better as expected, achieving 7.8\%--23.6\% scores with our set of models.}


\begin{table}[t]
\centering
\caption{Comparing the adequacy of model-generated and developer-written tests.}
\resizebox{\columnwidth}{!}{%
\renewcommand{\arraystretch}{1.2}
\begin{tabular}{lllrrrr}
\toprule
\multicolumn{1}{c}{\multirow{2}{*}{Method}} & \multicolumn{1}{c}{\multirow{2}{*}{Category}}                                     & \multicolumn{1}{c}{\multirow{2}{*}{Model}} & \multicolumn{1}{c}{\multirow{2}{*}{\# of Tests}} & \multicolumn{2}{c}{Adequacy}                                                                                                                                              & \multicolumn{1}{c}{\multirow{2}{*}{p-value}} \\ \cmidrule{5-6}
\multicolumn{1}{c}{}                        & \multicolumn{1}{c}{}                                                              & \multicolumn{1}{c}{}                       & \multicolumn{1}{c}{}                                 & \multicolumn{1}{c}{\begin{tabular}[c]{@{}c@{}}Model \\ Generated\end{tabular}} & \multicolumn{1}{c}{\begin{tabular}[c]{@{}c@{}}Developer \\ Written\end{tabular}} & \multicolumn{1}{c}{}                         \\ \midrule
\multirow{6}{*}{Zero-shot}                  & \multirow{3}{*}{\begin{tabular}[c]{@{}l@{}}Fail-to-Pass\\ by model\end{tabular}}  & Llama-3.1                                  & 51                                                   & 0.91                                                                                 & 0.95                                                                               & 0.03                                         \\
                                            &                                                                                   & Mistral-large                              & 57                                                   & 0.93                                                                                 & 0.96                                                                               & 0.04                                         \\
                                            &                                                                                   & GPT-4o                                     & 84                                                   & 0.92                                                                                 & 0.95                                                                               & 0.01                                         \\ \cline{2-7}
                                            & \multirow{3}{*}{\begin{tabular}[c]{@{}l@{}}Others by \\ model\end{tabular}}       & Llama-3.1                                  & 329                                                  & 0.55                                                                                 & 0.94                                                                               & \textless{}0.001                             \\
                                            &                                                                                   & Mistral-large                              & 340                                                  & 0.59                                                                                 & 0.94                                                                               & \textless{}0.001                             \\
                                            &                                                                                   & GPT-4o                                     & 313                                                  & 0.53                                                                                 & 0.93                                                                               & \textless{}0.001                             \\ \midrule
\multirow{6}{*}{Auto-TDD}                   & \multirow{3}{*}{\begin{tabular}[c]{@{}l@{}}Fail-to-Pass \\ by model\end{tabular}} & Llama-3.1                                  & 37                                                   & 0.95                                                                                 & 0.95                                                                               & 0.59                                         \\
                                            &                                                                                   & Mistral-large                              & 85                                                   & 0.97                                                                                 & 0.99                                                                               & 0.04                                         \\
                                            &                                                                                   & GPT-4o                                     & 106                                                  & 0.96                                                                                 & 0.98                                                                               & 0.02                                         \\ \cline{2-7}
                                            & \multirow{3}{*}{\begin{tabular}[c]{@{}l@{}}Others by \\ model\end{tabular}}       & Llama-3.1                                  & 337                                                  & 0.49                                                                                 & 0.94                                                                               & \textless{}0.001                             \\
                                            &                                                                                   & Mistral-large                              & 349                                                  & 0.53                                                                                 & 0.93                                                                               & \textless{}0.001                             \\
                                            &                                                                                   & GPT-4o                                     & 332                                                  & 0.52                                                                                 & 0.93                                                                               & \textless{}0.001       
                                            \\ \bottomrule
                                            \multicolumn{4}{l}{*p-value is computed using pairwise Wilcoxon signed-rank test.}
\end{tabular}
}
\label{tbl:coverage-ori-model}
\vspace{-15pt}
\end{table}

\subsection{RQ4: Test Adequacy}
\label{sec:result-rq2}

\cref{tbl:tdd-performance} incorporates coverage as part of the $\mathit{tddScore}$ column. \cref{tbl:coverage-ori-model} drills down deeper on coverage, and we have seen very similar statistics for both of the baseline and \atdd approaches. First, even human-written tests are not perfect and, in our benchmark, \emph{the final score achieved by golden tests $\hat{y}$ is} 0.94. Next, we evaluate how the model-generated tests stack up with respect to coverage. Note that, all of the golden tests are fail-to-pass, but all model-generated ones are not. So, we discuss them in two separate groups: fail-to-pass tests and other tests.
For all models and approaches, we found that model-generated fail-to-pass tests achieve 0.91--0.95 coverage, which is very close to the coverage achieved by developer-written tests. To make a fair comparison, we do not consider all the samples from human-generated tests. If model $M$ has 100 fail-to-pass tests, we collected coverage for exactly those 100 instances from the developer-written set and made the comparison. We also performed a non-parametric pairwise Wilcoxon signed rank test and failed to reject the null hypothesis at 99\% confidence interval.

To summarize, we observe that the coverage achieved by model-generated tests is slightly lower than that of human-written ones, but the difference is not statistically significant. On the other hand, if the tests are not fail-to-pass, they have much lower coverage compared to human-written ones. This indicates that it is quite difficult for tests to go from fail to pass without covering the deleted and added lines between between the old code and the patched code. 

\findings{4}{Model-generated fail-to-pass tests achieve similar coverage as the developer-written test (above 0.9). However, for other tests (e.g., fail-to-fail), the coverage is considerably low (below 0.6). This finding is independent of the models and approaches we used.}

\section{Discussion}
\label{discussion}







\subsubsection*{Relation between test adequacy and correctness}
Although we have seen higher coverage for fail-to-pass tests compared to other tests, coverage does not indicate correctness. We have seen 35\% of non-fail-to-pass tests with perfect coverage using GPT-4o based \atdd. This indicates that tests can achieve perfect coverage even without being fail-to-pass. 
However, we have successfully localized several issue-related functions in \atdd. 
If we can localize the position of the lines of code that would be deleted or updated to address the issue, we can use low coverage to discard candidates that are very unlikely to go from fail to pass because, for almost all fail-to-pass tests, the coverage is above 0.90.

\subsubsection*{Uniqueness of fail-to-pass tests generated by different models}
\begin{figure}[!t]
    \centering
    \includegraphics[trim=0 0cm 0 0cm, scale=0.48]{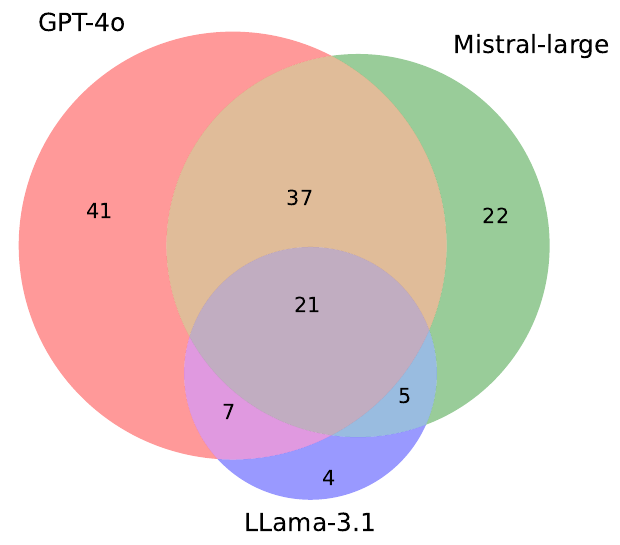}
\caption{Number of instances with fail-to-pass tests generated by different models.}
    \label{fig:venn}
    \vspace{-.5cm}
\end{figure}

\cref{fig:venn} shows the number of instances with fail-to-pass tests generated by different models. GPT-4o, Mistral, and \mbox{Llama-3.1} can uniquely generate fail-to-pass tests for 41, 22, and 4 instances, respectively. This means there are 67 instances uniquely solved by a single model. Additionally, fail-to-pass tests for 68 instances (37+21+7+5) can be generated by multiple models. Note that the fail-to-pass tests for a single instance generated by multiple models are not the same. Each model can generate completely different fail-to-pass tests. The union of all three models can generate fail-to-pass tests for a total of 135 out of 449 (30\%) instances altogether. This indicates that an ensemble approach could potentially increase the performance of \atdd, though with an increase in inferencing cost. We leave this for future research.

\section{Threats to Validity}
\label{sec:threat}
One major limitation of \tdd is that it is mined from 12 popular Python repositories, so findings may not apply to other programming languages and repositories.
We note that SWE-bench, despite having the same limitations, has been
impactful, and one of the findings in the SWE-bench paper was that
``difficulty does not correlate with issue resolution date'',
indicating that contamination problems (if any) are minor~\cite{jimenezswe}.
A limitation of \atdd is that it only considers one test file or generate one block of code.
In real life, test code can be spread across multiple files or blocks of code.
\atdd cannot generate tests that must be written across multiple files.
Despite that, \atdd exceeded the state-of-the-art performance,
so we leave further improvements to future work.
We use the Python \texttt{\small coverage} package for computing test coverage, but this package can fail for various reasons, such as permission issues, version incompatibility, or configuration problems. In \tdd, we computed coverage for all projects, including SymPy. However, upon manual validation, we found the coverage information for SymPy to be unreliable. Therefore, we removed coverage from the final metric for SymPy instances.
However, given that coverage for fail-to-pass tests was consistently above 0.9 and fewer than 15\% of instances came from SymPy, this likely makes $<1.5$\% difference for the results.




\section{Related Work}
\label{sec:relatedwork}

The introduction already discussed the most closely related work by Kang \etal~\cite{kang2023large}, Plein \etal~\cite{plein_et_al_2024}, and M{\"u}ndler \etal~\cite{mundler2024code}.
Other related works can broadly be categorized into (a)~benchmarks to evaluate the quality of code generated by various automated approaches, including LLMs, and (b)~evaluations of automated test generation capabilities.

\paragraph{Code-related benchmarks}
There has been a large body of works on creating benchmarks for code-related tasks. For instance, there are works on code translation~\cite{avatar, evalplus, pan2024lost}, code generation~\cite{humaneval, javabench}, code repair~\cite{repair1, repair2}, code summarization~\cite{ahmed2024automatic, ahmed2022few}, code review~\cite{hellendoorn_et_al_2021}, and issue fixing~\cite{jimenezswe, empirical1, empirical2, empirical3, empirical4, empirical5}. Among these works, the closest works are related to fixing GitHub issues. Recently, with the emergence of benchmark like SWE-bench~\cite{jimenezswe}, there has been a significant contribution in this direction. This includes works that enhance the SWE-Bench dataset by adding support for more programming languages~\cite{swejava} (arxiv), enabling dataset for multi-modal model by introducing the visual aspect of the issues~\cite{swemultimodal} (arxiv), and performing more rigorous evaluation by applying various approaches and models~\cite{sweplus} (arxiv); building agentic workflows to resolve the issues from SWE-Bench (\cite{sweagent3coder, sweagent4magis, sweagent2, sweagent5agentless, sweagent6masai, sweagent7pybench, autocoderover2}  (arxiv) and~\cite{autocoderover}); and evaluating techniques such as chain of thoughts~\cite{issuetot} (arxiv), understanding resolving issues~\cite{repounderstander} (arxiv), etc. 
Compared to these works, we focus on extending the capability of SWE-Bench to evaluate the correctness and adequacy of tests. 

Also, there are works that create datasets to evaluate the quality of tests generated by LLMs~\cite{testagent1, testeval, pytestbench}. However, when it comes to resolving specific use cases such as resolving GitHub issues,  two of the closest works are done by Jain et al.~\cite{testgeneval} (arxiv) and M{\"u}ndler et al.~\cite{mundler2024code}. Jain et al.~\cite{testgeneval} focus on creating a parallel dataset similar to SWE-Bench but for test generation. The objective is to evaluate the capability of LLMs in test generation and test completion given a body of code. Compared to that, our objective is slightly different. Our starting point is an issue description and we focus on evaluating the quality of the tests generated by LLMs on hidden code patches.
M{\"u}ndler et al.~\cite{mundler2024code} built SWT-Bench, a similar
dataset to ours, but with less rigorous filters, leading to more but
lower-quality instances than \tdd.
Due to the prohibitive cost of running the full SWT-Bench, M{\"u}ndler
et al.\ only experiment with a subset SWT-Bench Lite filtered to be
less demanding.
Also, the SWT-bench evaluation harness measures coverage in a more
round-about way than \tdd.


\paragraph{Evaluating test generation capability} 
There have been several works that attempted to use LLMs for test generations~\cite{wang2024software, tufano2020unit, vikram2023can, bareiss2022code,ryan2024code, schafer2023empirical, pan2024multi}. These works are mostly tied to the capability of LLMs to generate unit tests for Java, Python, and other programming languages. Also, there are works that create both benchmarks and evaluate the capability of LLMs in generating tests. Compared to that, our focus is on generating tests for GitHub patches and the setting of test-driven development.
\section{Conclusion}
\label{sec:conclusion}

This paper contributes \tdd, a challenging new benchmark for
test-generation directly from issue descriptions before anyone writes
the code to be tested.
\tdd is mined from real-world GitHub issues with strict filters and
evaluation metrics.
This paper also contributes \atdd, an LLM-based solution for \tdd that
outperforms the previous state-of-the-art for this problem.

\bibliographystyle{IEEEtran}
\bibliography{reference}


\end{document}